\documentclass[twocolumn,prb,aps,showpacs,amsmath,amssymb]{revtex4}

\usepackage{graphicx}
\usepackage{dcolumn}
\usepackage{bm}

\begin{document}




\title{Cellular Dynamical Mean Field Theory for the 1D Extended
Hubbard Model}

\author{C.~J.~Bolech}
\author{S.~S.~Kancharla}
\author{G.~Kotliar}

\affiliation{Center for Materials Theory,
             Serin Physics Laboratory,
             Rutgers University\\
             136 Frelinghuysen Road,
             Piscataway,
             New Jersey 08854-8019,
             USA}

\date{\today}

\begin{abstract}
We explore the use of exact diagonalization methods for solving the
self consistent equations of the cellular dynamical mean field theory
(CDMFT) for the one dimensional regular and extended Hubbard models.
We investigate the nature of the Mott transition and convergence of
the method as a function of cluster size as well as the optimal
allocation of computational resources between bath and
``cluster-impurity'' sites, with a view to develop a renormalization
group method in higher dimensions. We assess the performance of the
method by comparing results for the Green's functions in both the spin
density wave (SDW) and charge density wave (CDW) phases with accurate
density matrix renormalization group (DMRG) calculations.

\end{abstract}



\maketitle


\section{Introduction}
The dynamical mean field theory (DMFT)\cite{rmp} has been successful
in accessing certain aspects of the non-perturbative phenomena in
strongly correlated electron systems. For example, it has given new
insights and predicted various qualitative trends in the
redistribution of spectral weight as temperature, pressure and doping
are varied in optical and photomission experiments, near the Mott
transition.

Single site DMFT extracts physical quantities such as the self energy
from a local impurity model with a self consistent bath (also refered to
as LISA for local impurity self consistent approximation) and by
construction this approach misses the effects of short range
correlations, such as the $k$-dependence of the self energy, which are
bound to become important at low temperatures. It also does not allow
the treatment of phases such as d-wave superconductivity, and hence
there have been significant efforts to extend the DMFT
methodology\cite{rmp,dca1,dca2,lichtenstein,cdmft,pankov,chaindmft,ping}.

Many of the elements of the DMFT method can be traced back to the
coherent potential approximation (CPA)\cite{leath} which has been very
successful in disordered systems. In this context, the search for
extensions of the CPA was fraught with many difficulties, and
extensions of DMFT should be scrutinized carefully in this light. The
DMFT equations, while having a similar spirit to those of CPA for
determining the effective medium, are considerably more complex
because they involve the solution of a quantum impurity model, which
plays the role of the effecive Hamiltonian for the local degrees of
freedom treated in the DMFT. To analyse these equations new concepts
and numerous techniques have been developed over the last
decade. Their incorporation into cluster methods is promising but
requires substantial new work. This paper is a contribution in this
direction.
In this paper we focus on one cluster extension of DMFT, the Cellular
Dynamical Mean Field Theory (CDMFT)\cite{cdmft,biroli}.
The aim of this paper is to  investigate how the exact
diagonalization aproach which was so successful in the context of
the single site DMFT\cite{krauth,rozenberg,si}, can
be used in the CDMFT context.

We solve the self consistent cluster equations for the one dimensional
Hubbard and extended Hubbard models using exact diagonalization of the
effective underlying cluster impurity. We choose these models because
of their generic nature, as well as the fact that DMFT, which becomes
exact in the limit of infinite coordination number\cite{metzner},
faces the worst case scenario in one dimension. Furthermore in one
dimension fairly reliable computations of static as well as dynamic
quantities can be made using the Density Matrix Renormalization Group
(DMRG)\cite{white,hallberg,kuhner} approach to carry out a comparative
study.

The exact diagonalization approach, within the single site DMFT
context, lead to the development of a technique inspired by
renormalization group ideas which resulted in the first
quantitative study of the critical properties near the Mott
transition\cite{moeller}. With a view towards the future, we 
discuss the possibility of combining renormalization group ideas and
CDMFT to develop a new numerical method in the spirit of the DMRG.

The Extended Hubbard Model (EHM) is defined by the following Hamiltonian:
\begin{eqnarray}
H&=&-t\sum_{j,\sigma}\left( c_{j+1\sigma}^{\dagger }c_{j\sigma}+ {\rm h.c.} \right)
+ U\sum_{j} n_{j\uparrow} n_{j\downarrow} \nonumber \\
  && + V\sum_{j} n_{j}n_{j+1}-\mu \sum_{j} n_{j}
\label{exthubbard}
\end{eqnarray}
The first term corresponding to hopping between nearest neighbor sites
and the second term to the onsite Coulomb repulsion provide the
competition between itineracy and localization in the regular Hubbard
model. The third term represents Coulomb repulsion between electrons
occupying nearest neighbor sites. The Hamiltonian as written above
with $\mu=U/2+2V$ guarantees an insulating ground state with a
filling of one electron per site.

In the following section we introduce and discuss the CDMFT self
consistency equations and show how they can be generalized to treat
non-local Coulomb interactions even in the presence of broken
symmetries. In the next and subsequent sections we present results for
the regular Hubbard and the extended Hubbard model respectively. This
is followed by the conclusions and an outlook for further work.

\section{The CDMFT Method}
The CDMFT\cite{cdmft} method is a straightforward generalization of
the single site DMFT, in which local degrees of freedom within a
cluster are treated exactly and those outside the cluster are replaced
by a bath of non-interacting electrons determined self
consistently. This approach, formulated with realistic studies in
mind, deals in principle with overlapping cells and with clusters
which are not necessarily defined in real space and could even be
defined by non-orthogonal orbitals. In this paper, we ignore these
complications and work with real space clusters using an orthogonal
basis. It is worth pointing out, that if we apply the approach to
non-interacting disordered alloys, the self consistency condition of
CDMFT becomes identical to that used in the molecular CPA
approximation of Ducastelle\cite{ducastelle}. This is in the same way
as the self consistency condition in single site DMFT becomes
identical to the single site CPA and the two impurity approximation
proposed by Ingersent and Schiller\cite{ingersent} and by Georges and
Kotliar\cite{rmp}, becomes identical to the pair CPA. However,
unlike the single site case, where the transition from real space to
momentum space is unambiguous, one has to be careful in interpreting
the results of the CDMFT equations in $k$-space. This is because the
goal of CDMFT is to obtain the best possible estimates of {\it local}
quantities that live within a cluster. For this purpose, it introduces
a cluster self energy. Long distance properties such as the ones
contained in the lattice Green's function and the lattice self energy
are then {\it inferred} from the cluster Green's function or the cluster
self energy while maintaining causality\cite{cdmft}.

\subsection{The Cavity Construction}
The most economical way to arrive at the concepts of a dynamical mean
field theory is via the cavity construction, which stresses the point
made above, that the focus of the method is in extracting local
quantities. The original lattice is divided into equal clusters of
size $N_{c}$. Integrating out the degrees of freedom external to a
chosen cluster (labelled $0$), we can formally write an effective
action which would allow the computation of local quantities as
\begin{equation}
\frac{1}{Z_{\rm eff}}e^{-S_{\rm eff}\left[ c_{0\mu \sigma }^{\dagger },c_{0\mu
\sigma }\right] }\equiv \frac{1}{Z}\int \prod_{j\neq 0,\mu \sigma
}{\cal D}c_{j\mu \sigma }^{\dagger }{\cal D}c_{j\mu \sigma }e^{-S}.
\end{equation}
Here $j$ labels individual clusters and $\mu$ is an index labelling
sites within each cluster. We can split the original action into three
parts,
\begin{equation}
S=S^{(0)}+S_{0}+\Delta S,
\end{equation}
where $S^{(0)}$ includes terms outside the chosen cluster (the full
action with the cluster replaced by a cavity), $S_{0}$ comprises of
terms solely within the cluster and finally, $\Delta S$ includes those
terms that couple the cluster with its environment. Explicitly, in the
case of the EHM,
\begin{eqnarray}
S_{0}&=&\int_{0}^{\beta }d\tau
\sum_{\mu \nu ,\sigma \rho }c_{0\mu \sigma }^{\dagger }(\delta _{\mu
\nu }\delta _{\sigma \rho }\partial _{\tau }+{E}_{\mu \nu ,\sigma \rho
}^{00})c_{0\nu \rho } \nonumber \\
&& + \sum_{\mu }Un_{0\mu \uparrow }n_{0\mu \downarrow
}+V\sum_{\left\langle \mu ,\nu \right\rangle }n_{0\mu }n_{0\nu}
\end{eqnarray}
\begin{eqnarray}
\Delta S &=&\int_{0}^{\beta } d\tau
\sum_{\left\langle j\mu ,0\nu \right\rangle ,\sigma \rho }{E}_{\mu \nu
,\sigma \rho }^{j0}c_{j\mu \sigma }^{\dagger }c_{0\nu \rho }+{E}_{\nu
\mu ,\rho \sigma }^{0j}c_{0\nu \rho }^{\dagger }c_{j\mu \sigma
} \nonumber \\
&& + V\sum_{\left\langle j\mu ,0\nu \right\rangle }n_{j\mu }n_{0\nu}
\end{eqnarray}
Here ${\hat E}^{00}$ includes the hopping matrix as well as the
chemical potential within the zeroeth cluster and $\left\langle
j\mu,0\nu\right\rangle$ denotes all intercluster nearest neighbors. It
is to be noted that, while the onsite Coulomb interaction $U$
contributes only to $S_{0}$, a nonlocal interaction such as the
nearest neighbor Coulomb repulsion $V$ contributes to both $S_{0}$ as
well as $\Delta S$. This cavity construction is so far merely a
relabelling of terms in the original action and is exact. However
approximations need to be made to actually access the local properties
within the cluster. We approximate the effective action of the cluster
by keeping only the renormalization of the quadratic terms obtained
after integrating out the degrees of freedom of the surrounding
environment\cite{rmp}. Notice that this approximation violates
the translational symmetry of the original lattice. But this is not a
problem, since the spirit of the approach is to estimate local
quantities. The Gaussian approximation for the effects of the
environment seen by the cluster becomes exact in the limit of large
coordination\cite{rmp} and therefore, our test case in one dimension
constitutes the worst case scenario. The effective action for the
cluster can then be approximated as,
\begin{eqnarray}
S_{\rm eff}\left[ c_{0\mu \sigma }^{\dagger },c_{0\mu \sigma }\right] &=&
\int_{0}^{\beta }d\tau \sum_{\mu \nu ,\sigma \rho }c_{0\mu \sigma
}^{\dagger }\left[ \hat{\cal G}_{0}^{-1} \right]
_{\mu \nu ,\sigma \rho }c_{0\nu \rho } \nonumber \\ &+& \sum_{\mu
}Un_{0\mu \uparrow }n_{0\mu \downarrow }+V\sum_{\left\langle \mu ,\nu
\right\rangle }n_{0\mu } n_{0\nu }.
\end{eqnarray}
Here, the time dependent Weiss field $\hat{\cal G}_{0}^{-1}$ is now a
matrix in the cluster variables and is a functional of the EHM Green's
function with the cluster replaced by a cavity. Suppressing spin
indices, it can be written on the Matsubara axis as
\begin{equation}
\begin{split}
\left[ \hat{\cal G}_{0}^{-1}\left( i\omega _{n}\right) \right]_{\mu\nu} &= i\omega
_{n}\delta _{\mu \nu }-{E}_{\mu \nu}^{00} -V\,\delta _{\mu \nu
}\sum_{\left\langle j\mu ^{\prime },0\nu \right\rangle }\left\langle
n_{j\mu ^{\prime}}\right\rangle _{\left( 0\right) } \nonumber \\ &-
\sum_{\left\langle i\mu ^{\prime },0\mu \right\rangle}\sum_{\left\langle j\nu ^{\prime },0\nu \right\rangle} {E}_{\mu \mu
^{\prime }}^{0i}G_{i\mu ^{\prime },j\nu ^{\prime }}^{\left( 0\right)
}\left( i\omega_{n}\right) {E}_{\nu ^{\prime }\nu }^{j0}.
\end{split}
\end{equation}
Relating the cavity Green's function, $G^{(0)}$, to the full Green's
function of the EHM, the following self consistency condition to
determine the Weiss field can be obtained:
\begin{equation}
\hat{\cal G}_{0}^{-1}(i\omega _{n}) = \hat{G}_{loc}^{-1}(i\omega _{n})+{\hat \Sigma_c}(i\omega _{n}),
\label{selfcon}
\end{equation}
where the local Green's function on the cluster in one dimension is
defined as (with an obvious generalization to higher dimensions)
\begin{equation}
\hat{G}_{loc}(i\omega _{n})=\int_{-\pi /N_{c}}^{\pi /N_{c}}\frac{1}{i\omega
_{n}+\mu -{\hat \Sigma_c}(i\omega _{n})-\hat{t}(k)}\frac{dk}{2\pi /N_{c}}.
\end{equation}
$\hat \Sigma_c$ is the cluster self energy as obtained from $S_{\rm
eff}$, $\hat{t}(k)$ is the Fourier transform of the cluster hopping
matrix of the EHM and $k$ is a vector in the reduced Brillouin zone of
the superlattice $(-\frac{\pi }{N_{c}},\frac{\pi
}{N_{c}}]$. Explicitly for a one dimensional model with only nearest
neighbor hopping,
\begin{eqnarray}
{t}^{\mu \nu }(k)&=&-t[\delta _{\mu -\nu \pm 1}+e^{-ikN_{c}}\delta _{N_{c}+\mu
-\nu \pm 1}\nonumber \\ &&+e^{ikN_{c}}\delta _{-N_{c}+\mu -\nu \pm 1}]
\end{eqnarray}

\subsection{A functional interpretation}
A second approach to derive the above approximation is via the
construction of a functional of the quantities of interest, say the
Green's function restricted to a local cluster and its supercell
translations, such that the stationary point of the functional yields
those quantities\cite{chitra}. In principle such a functional can be
constructed order by order in a perturbative expansion in the hopping
or the interactions while constraining the Green's function to a fixed
value. In practice, sensible approximations to the exact functional
are contructed by starting with the full Baym-Kadanoff functional
$\Phi[G]$ and restricting its argument from the full Green's function to
the Green's function defined only inside the cluster and its periodic
supercell repetitions. This constraint is realized by the cluster self
energy which plays the role of the Lagrange
multiplier\cite{fukuda,chitra,kotliar}. Differentiation of this
functional gives the CDMFT equations.  Viewed as an approximation to
the Baym-Kadanoff functional for the full lattice, this procedure may
seem a bad approximation\cite{gonzalez}. However, this construction of
the functional should be viewed as an approximation to the exact
effective action for local quantities and CDMFT uses it precisely for
such a purpose.

The functional approach is useful because it clarifies the level of
approximation used in treating the broken symmetry state. In the case
that we consider in this paper, all the Hartree graphs are added to
the functional (see Eq.~\ref{hartree}). The
E-DMFT\cite{pankov,ping,chitra,smith} includes additional graphs to the
Hartree terms, but the quality of the results presented below show
that they would make a very small contribution.

\subsection{An Exact Diagonalization Algorithm}
For the purpose of practical
implementation of the algorithm it is convenient to consider the
cluster effective action to arise from a generalized impurity
Hamiltonian,
\begin{eqnarray}
H_{imp}&=&\sum_{\mu \nu \sigma }{E}_{\mu \nu }c_{\mu \sigma }^{\dagger
}c_{\nu \sigma }+U\sum_{\mu }n_{\mu \uparrow }n_{\mu
\downarrow }\nonumber \\&&
+\sum_{\mu \nu }{V}_{\mu \nu }n_{\mu }n_{\nu }+\sum_{k\sigma }\epsilon
_{k\sigma }a_{k\sigma }^{\dagger }a_{k\sigma }\nonumber \\&&+\sum_{k\sigma ,\mu
}(V^h_{k\sigma ,\mu }a_{k\sigma }^{\dagger }c_{\mu \sigma }+V^{h*}_{k\sigma
\mu }c_{\mu \sigma }^{\dagger }a_{k\sigma })
\end{eqnarray}
Here, $\epsilon_{k\sigma}$ represents the dispersion of an auxiliary
non-interacting bath and $V^h_{k\sigma,\mu }$ is the hybridization
matrix between the bath and the impurity cluster. The second and third
terms include all the intra-cluster interaction terms. In
$\hat{E}$ we lump together the cluster hopping matrix and
chemical potential, as well as the Hartree terms arising from
non-local interactions between clusters. For the EHM
\begin{equation}
{E}_{\rho \zeta}= {E}_{\rho \zeta}^{00}+V\,\delta _{\rho \zeta
}\sum_{\left\langle j\rho ^{\prime },0\zeta \right\rangle }\left\langle
n_{j\rho ^{\prime}}\right\rangle,
\label{hartree}
\end{equation}
where $\left\langle n_{j\rho ^{\prime}}\right\rangle$ is computed by
using the translational invariance of the system at the level of the
superlattice.

In terms of the impurity model the parametrization of the Weiss field
function is readily obtained as,
\begin{equation}
\hat{\cal G}_{0}^{-1}(i\omega_{n})=i\omega_{n}-\hat{E}-\hat{\Delta}(i\omega_{n});
\end{equation}
where the hybridization function is given by
\begin{equation}
\Delta_{\mu \nu ,\sigma
}(i\omega_n)[\epsilon_{k\sigma},V^h_{k\sigma,\delta}]= \sum_{k}
\frac{V^{h*}_{k\sigma ,\mu}V^h_{k\sigma ,\nu
}}{i\omega_{n}-\epsilon_{k\sigma }}.
\end{equation}
Thus the algorithm involves determining the bath parameters
$\epsilon_{k\sigma}$ and $V^h_{k\sigma,\mu }$ self consistently
subject to the condition given by Eq.~\ref{selfcon}. Our
implementation relies on the solution of the impurity model using
exact diagonalization\cite{krauth}. Starting with an inital guess for
the bath parameters we obtain the cluster self energy $\hat \Sigma_c$,
and using the self consistency condition determine a new Weiss field,
$\hat{\cal G}_{0}^{-1}(i\omega_{n})^{\rm new}$. In turn, this defines
a new hybridization function $\hat{\Delta}(i\omega_{n})^{\rm new}$. To
close the self consistency loop we project onto a finite subspace of
bath size $N_b$. This projection is carried out using a conjugate
gradient minimization of the following distance function,
\begin{equation}
D=\frac{1}{\left( n_{\rm max }+1\right) N_{b}N_{c}}\sum_{n=0}^{n_{\rm max
}}\left\| \hat{\Delta}(i\omega _{n})^{\rm{new}}-\hat{\Delta}(i\omega
_{n})^{N_{b}}\right\|
\end{equation}
where $n_{\rm max}$ is the number of grid points on the Matsubara
axis. Although we study the CDMFT equations only at zero temperature,
the self consistency equations are solved on the Matsubara axis.




\section{Results for the Hubbard Model ($V=0$)}
In order to highlight the differences between the single impurity
(LISA) and cluster dynamical mean field schemes we present results
obtained for the local spectral gap as a function of the onsite
Coulomb repulsion $U$. To keep the computational cost similar, we
initially fix the total number of sites in the cluster and bath,
$N_s=N_c+N_b=6$, across schemes. The exact result for the gap is known
from Bethe ansatz and is given by\cite{ovi}
\begin{equation}
\Delta (U)= \frac{16t^2}{U}\int_{1}^{\infty} \frac{\sqrt{y^2-1}}{\sinh(2\pi ty/U)}dy
\end{equation}
\begin{figure}[ht]
\includegraphics[width=0.4\textwidth]{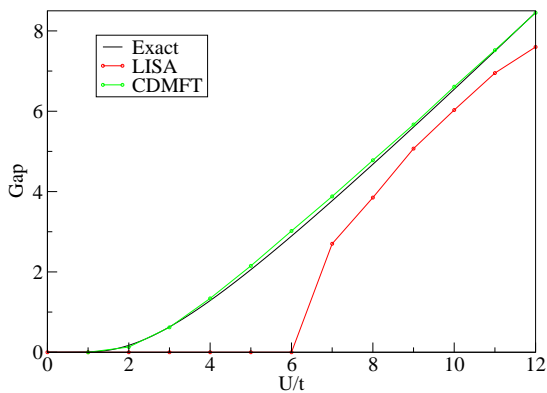}
\caption{\label{gaps}Spectral gap as a function of U/t for the half
filled Hubbard model.}
\end{figure}

Fig.~\ref{gaps} shows that there is a characteristically different
behavior between the single impurity ($N_c=1$) and cluster schemes
($N_c=2$).  In the single impurity case, for $U \agt 7.5t$ the gap
follows the exact result, whereas for $U \alt 7t$ it reduces to values
much smaller than the exact gap and approaches zero. For the range
$7t\alt U \alt 7.5t$ we observe a coexistence of the gapless and
gapful phases. This transition from an insulating to a metallic phase
is an artifact of the mean field approach which incorporates the
physics of higher dimensions where the Mott transition is indeed
present. On the other hand, CDMFT compares excellently to the exact
gap and an insulating solution exists through all finite values of
$U$, in agreement with the well known physics of the one dimensional
Hubbard model. We measure the half spectral gap as the point where the
strength of the lowest frequency pole falls to $9\%$ of its peak
height. This percentage is arrived at by requiring that the gap at
strong coupling be fixed at the exact value.  We also find, as
expected from the cavity construction and noted
recently\cite{gonzalez}, that in the case of CDMFT, the bath only
couples to sites on the boundary of the cluster, whereas for other
cluster methods such as the dynamical cluster approximation (DCA) all
the sites in the cluster are equivalently coupled to the bath. The
self consistent procedure is robust and generates these
two different kinds of solutions regardless of the nature of the
starting guess.

\begin{figure}[ht]
\includegraphics[width=0.45\textwidth]{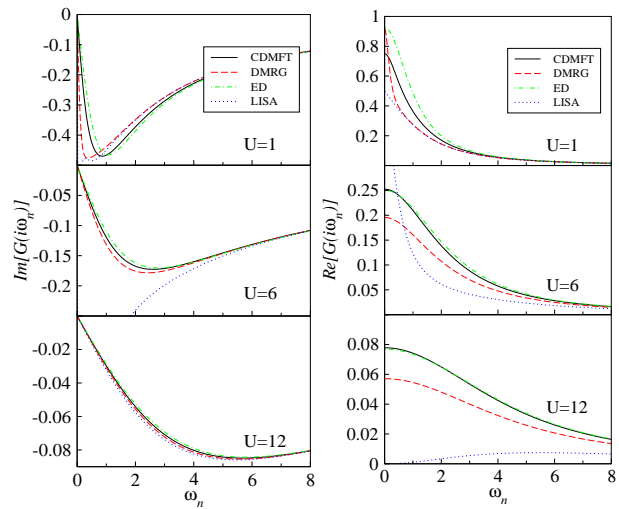}
\caption{\label{g11g12}Imaginary part of the onsite Green's function
(left column) and real part of the nearest neighbor Green's function
(right column) on the Matsubara axis for different values of
$U$ (in units of $t$).}
\end{figure}

\begin{figure*}[ht]
\includegraphics[width=0.95\textwidth]{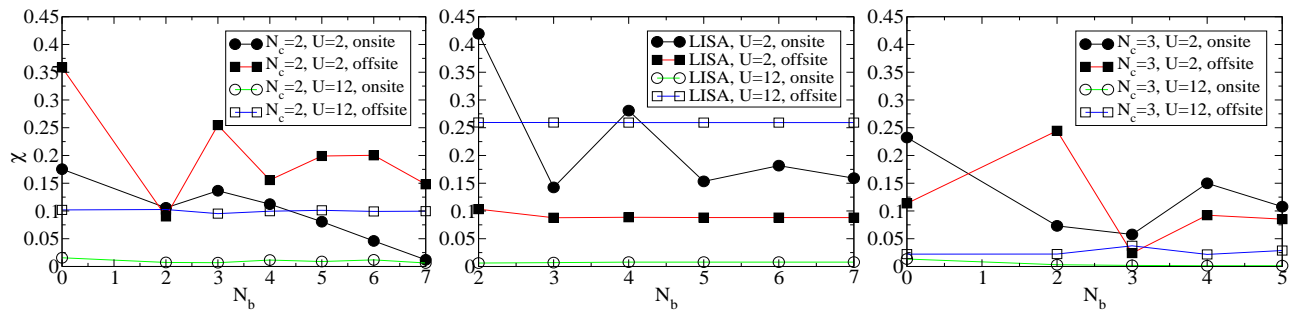}
\caption{\label{dist} Deviation of LISA and CDMFT Green's
functions from the DMRG results as a function of the bath size $N_b$.}
\end{figure*}

To further illustrate the performance of the CDMFT method, in
Fig.~\ref{g11g12} we compare the onsite as well as nearest neighbor
Green's functions on the Matsubara axis to those we obtain using the
DMRG method. The calculation of dynamical correlation functions within
the DMRG method is carried out by using the finite size
algorithm\cite{white} combined with the Lanczos vector
method\cite{hallberg,kuhner}. The results shown in the figures are
those for chains with 18 sites and open boundary conditions. We verify
that the results are close to those found by even longer chains and
hence representative of the system in the thermodynamic limit. To
highlight the role of the bath within the CDMFT method, we also
overlay results obtained from exact diagonalization (ED) of a cluster
of size $N_c=2$, but without the bath. As $U$ decreases we expect the
difference between onsite Green's functions from the isolated cluster
(ED) and the DMRG to increase, since the physics becomes less local.
The contribution of the bath in CDMFT therefore becomes more
significant with decreasing $U$ and shows a systematic improvement
with bath size as will be discussed later. Throughout the range of $U$
we consider, the imaginary part of the onsite Green's function
computed using CDMFT matches remarkably well with the DMRG curve. On
the other hand a comparison for the real part of the nearest neighbour
Green's function shows that the self consistent CDMFT solution
achieves only a marginal improvement over the ED calculation for small
$U$ and almost none for large $U$. It can be clearly seen that the
single impurity approach (LISA) performs poorly for the onsite Green's
function for all $U$, except when $U$ is large. Likewise, the nearest
neighbor Green's function computed using LISA agrees poorly with the
exact result throughout the entire range in $U$.


An important question to address is the issue of convergence and the
nature of the self consistent solution as a function of the size of
the effective impurity cluster and bath allotted in each case. In
order to find a satisfactory self consistent solution, we find that
the size of the Hilbert space allotted to the bath has to be at least
comparable to that alloted to the cluster itself. For instance,
clusters of size $N_c=2$ require a minimum bath of size $N_b=4$ for
good convergence. Even and odd clusters show qualitatively different
types of solutions, as we discuss further below. In Fig.~\ref{dist}
we plot $\chi$, a measure of the deviation of the LISA and CDMFT
solutions for the onsite and nearest neighbor Green's function from
accurate results obtained using DMRG, as a function of the bath
size $N_b$. This measure $\chi$, is defined as the integrated absolute
difference between the CDMFT (or LISA) and DMRG Green's functions.

Let us first discuss the results for the strong coupling case
($U=12t$). The onsite Green's function computed using LISA compares
excellently with DMRG and this continues to be so with the CDMFT
method. There is practically no dependence of $\chi$ on the bath size.
For the offsite Green's function, $\chi$ shows a systematic reduction
with increasing cluster size, but remains fairly independent of the
bath size.

For weak coupling, both LISA and CDMFT become exact. The toughest case
is when $U$ is of the order of the bandwidth; so in the following we
discuss the results for $U=2t$.  For even clusters, we find that
$\chi$ for the onsite Green's function shows a systematic decrease
with increasing bath size once a minimum bath size is reached. On the
other hand, $\chi$ for the offsite Green's function shows no definite
trend with increasing bath size.

\begin{figure}[ht]
\includegraphics[width=0.45\textwidth]{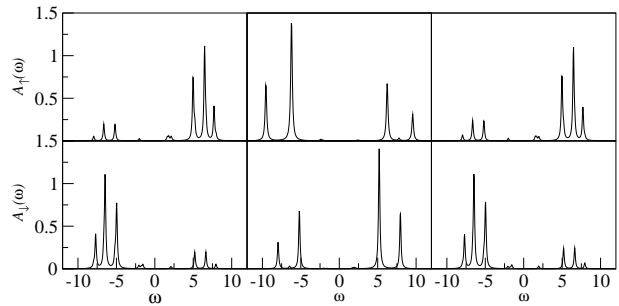}
\caption{\label{sdw} From left to right we show the local spectral
functions for sites one through three in a cluster of size
$N_c=3$. The bath size is fixed at $N_b=6$.}
\end{figure}


Contrary to the expectations for a mean field solution, even cluster
sizes ($N_c=2,4$) show no explicit local spin symmetry
breaking. Odd clusters ($N_c=3,5$) show similar behavior for small
bath sizes upto $N_b=N_c$. For larger bath sizes, $N_b\ge N_c$, and
large enough $U$, the behavior of the local spectral function is spin
dependent as can be seen from Fig.~\ref{sdw} for $U=12t$. Odd clusters
with broken symmetry are clearly inconsistent with cluster periodicty
and consequently show poorer convergence. Thus, for all even clusters
and odd clusters with small enough bath sizes, the one dimensional
character of the problem is dominant over the role of the bath and
prevents the self consistency from showing explicit spin symmetry
breaking. On the other hand, odd clusters (we were able to test only
the case $N_c=3$) with large $U$ and a sufficiently large bath show a
symmetry broken solution consistent with the mean field approach.


The effect of the bath for $N_c=3$ has an even more dramatic
consequence at smaller $U$. As opposed to even clusters which
correctly see only an insulating solution in one dimension, we see
both a metallic and an insulating solution for $N_b\ge5$ and
$U\le3t$. In the region of coexistence, the metallic solution shows a
better convergence than the insulating one. Thus, the improvement
gained in going from LISA to $N_c=3$ is that the region of coexistence
moves to smaller $U$'s and the insulating solution is present all the
way upto $U=0$. Small baths, due to the absence of enough degrees of
freedom to impose a mean field character to the solution, show an
insulating state throughout the range in $U$. Beyond $N_c=3$ for large
enough odd clusters we expect to see only an insulating solution for
all $U>0$.

\section{Results for the EHM ($V \neq 0$)}
\begin{figure}[ht]
\includegraphics[width=0.45\textwidth]{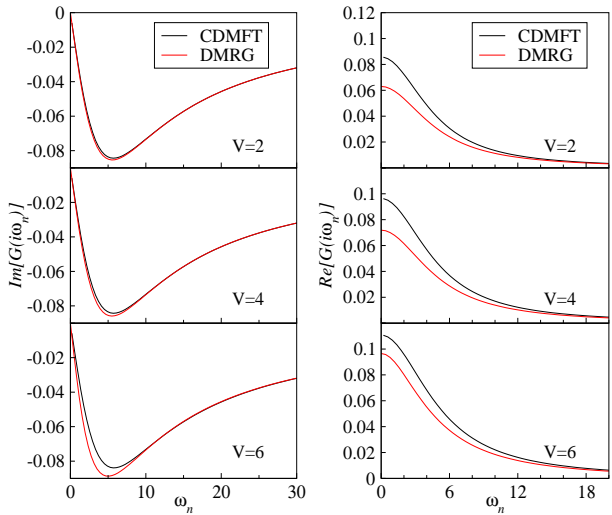}
\caption{\label{g11g12V}Imaginary part of the onsite Green's function
(left column) and real part of the nearest neighbor Green's function
(right column) on the Matsubara axis for different values of $V$ in
the SDW phase.}
\end{figure}
\begin{figure}[ht]
\includegraphics[width=0.4\textwidth]{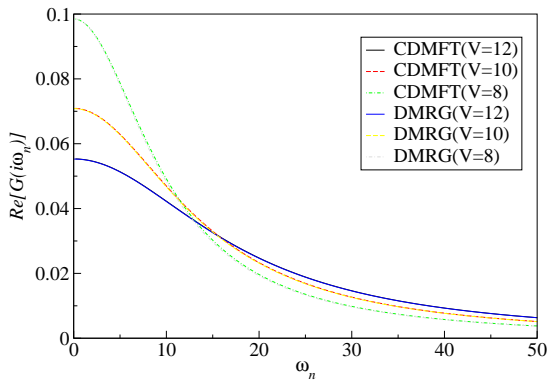}
\caption{\label{ReCDWg11}Real part of the onsite Green's function in
the CDW phase.}
\end{figure}
\begin{figure}[ht]
\includegraphics[width=0.4\textwidth]{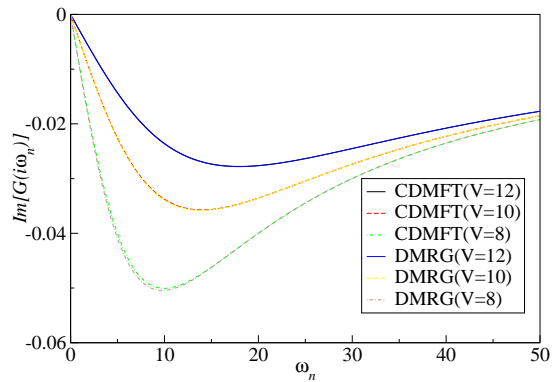}
\caption{\label{ImCDWg11}Imaginary part of the onsite Green's function
in the CDW phase.}
\end{figure}
\begin{figure}[ht]
\includegraphics[width=0.4\textwidth]{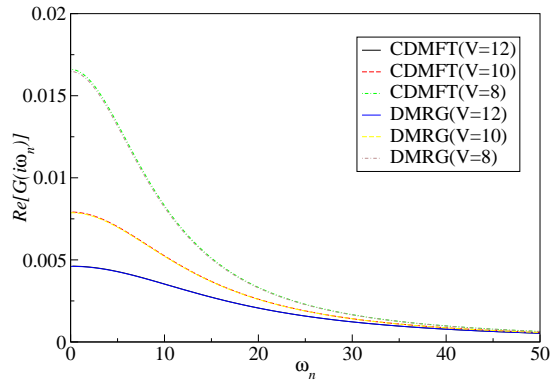}
\caption{\label{ReCDWg12}Real part of the nearest neighbor Green's
function in the CDW phase.}
\end{figure}
In this section we study the performance of cluster mean field methods
as applied to the Extended Hubbard model. This model constitutes a
natural test case since the non-local nearest neighbor Coulomb
repulsion among different sites within the cluster can be accounted
for exactly.

At half filling, for large enough $U$ (strong coupling), the system
goes through a first order transition from a spin density wave (SDW)
to a charge density wave (CDW) ordered state at roughly
$V=U/2$\cite{nakamura,ours,essler}. We work in this regime, by fixing
$U=12t$ and scanning through the nearest neighbor Coulomb repulsion
$V$. The CDMFT method shows a clear signature of this transition with
a cluster as small as $N_c=2$.  In Fig.~\ref{g11g12V} we report a
comparison of the imaginary part of the onsite Green's function and
the real part of the near neighbor Green's function in the SDW
phase. As seen in the $V=0$ case, the comparison with DMRG is better
for the former than the latter. We note that with increasing $V$ the
offsite Green's function compares better with the DMRG
result. Figs.~\ref{ReCDWg11},\ref{ImCDWg11} and \ref{ReCDWg12} compare
the Green's functions within the CDW phase. Across the range in $V$,
the agreement with DMRG is so remarkable that only one set of curves
are discernable within the figures. In particular, Fig.~\ref{ReCDWg11}
shows the real part of the onsite Green's function which is zero in
the SDW phase but acquires a non-zero value in the CDW phase. This
indicates a breaking of local particle-hole symmetry of the system
that is now conserved only on the average every two sites. Notice how
the curves for different values of $V$ cross, defining a characteristic
scale approximately equal to $U$. The versatility of CDMFT in being
able to treat the ordered phase stems from the fact that it only
assumes supercell periodicity and consequently allows for complicated
ordered states within a cell. This is in contrast to the DCA method
that assumes full periodicity of the lattice, and therefore requires a
fine discretization in momentum space to adequately capture the
discontinuities in the Brillouin zone that appear with the emergence
of short range ordered phases.

\section{Conclusions and Outlook}
\begin{figure}[ht]
\includegraphics[width=0.4\textwidth]{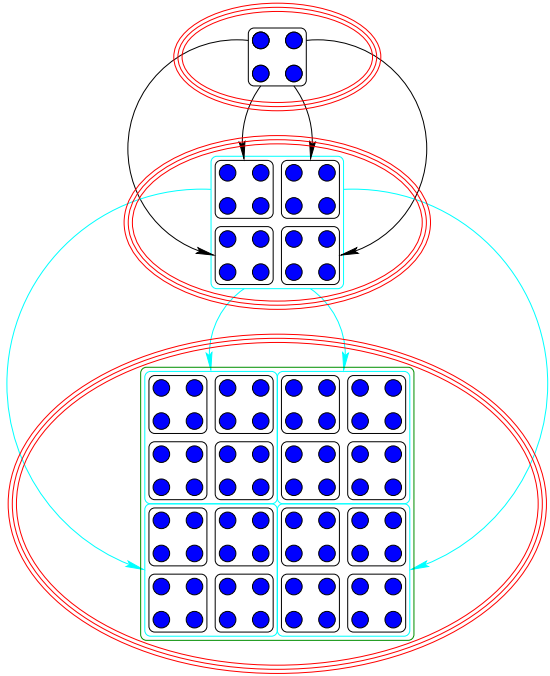}
\caption{\label{cdmftrg}Schematic representation of an RG procedure
intended to reach large cluster sizes.}
\end{figure}
We have shown that the CDMFT method produces remarkably good results
when tested in one dimensional systems. This is very encouraging,
since one would expect mean field methods to perform even better as
the dimensionality increases. We saw that even the smallest possible
cluster size ($N_c=2$) along with a self consistent bath of modest
size allows for an accurate determination of onsite correlations,
while the offsite quantities improve systematically with increasing
cluster size. Further, the success of our extension of the CDMFT
equations to describe the discrete broken symmetry of the CDW order
clearly demonstrates the power of the method to incorporate short
range correlations.

Following the example of the EHM, with the simplest kind of short
range correlation, it should be possible to treat the physics of more
complicated ordered phases as well as unit cells of realistic
materials within the cellular mean field approach. The main obstacle
in this program is the computational effort that goes into solving the
self consistent equations for the large clusters required by real
world applications. This calls for a systematic approach that blends
together cluster dynamical mean field and renormalization group
ideas\cite{wilson,white} to select the most relevant degrees of
freedom to represent a cluster.

We suggest that the CDMFT approach can be used iteratively in complete
analogy with the density matrix renormalization group approach. It is
useful to compare the similarities and differences between these two
methods. In both of them, one improves on the exact diagonalization of
a small finite size system by embedding the small system in a larger
one; an embedding which can be described by a reduced density
matrix. Within DMRG the density matrix is determined by an exact
diagonalization of the larger system containing the subsystem of
interest. In CDMFT, one can approximate the density matrix by modelling
the enviroment with a Gaussian Weiss field.

The CDMFT assumption of a self consistent gaussian bath as the
environment, that one can integrate out exactly in order to define a
reduced density matrix for the cluster, becomes more and more accurate
as the dimensionality increases. This choice of bath is optimum in a
dynamical mean field sense (optimized for the computation of the local
one particle Green's function of the cluster) and sidesteps the
conventional DMRG procedure for building the reduced density matrix
out of a few target states. Further, the DMRG prescription\cite{white}
of selecting the states for which the reduced density matrix has the
largest eigenvalues allows for a truncation of the cluster Hilbert
space while retaining the relevant physical information.  In
Fig.~\ref{cdmftrg} we indicate schematically how such a method would
proceed by doing CDMFT on small clusters, truncating their Hilbert
space using the DMRG prescription and finally using them as building
blocks of even larger clusters. This procedure can be repeated until
either a desired cluster size or convergence in a certain observable
of interest is reached. We believe this scheme should open new vistas
for numerical renormalization group calculations of realistic systems
in two and three dimensions.

\acknowledgments This work was supported by the NSF, under grant
DMR-0096462 and by the Rutgers Center for Materials Theory. Useful
discussions with G. Biroli, G. Palsson and S. Pankov are gratefully
acknowledged.

\appendix*

\section{Determination of the Cluster Self Energy}
The first step in the CDMFT iterative scheme is
to compute the cluster self energy $\hat \Sigma_c$ for an initial
guess of the bath parameters. This can be done using
Eq.~\ref{selfcon} by subtracting the inverses of the numerically
determined cluster Green's function ($\hat G_{imp}(i\omega_n)$)
from the exactly known Weiss field. Although this is accurate
enough for most purposes, when the results are small, it can
introduce numerical errors. To deal with this, an alternative
procedure was proposed in the context of the single impurity
Anderson model\cite{bulla}. The idea is to isolate contributions
to the self energy coming purely from the hybridization and the
interactions. The terms arising from the interactions can be
written as ratios of correlation functions of composite operators
leading to a more stable numerical procedure. We generalize this
procedure to the cluster Anderson impurity model to write the
self energy in the form
\begin{eqnarray}
{\hat \Sigma}_c(i\omega_n)={\hat \Sigma}_{U}(i\omega_n)+{\hat
\Sigma}_{V}(i\omega_n)
\end{eqnarray}
where
\begin{eqnarray}
\Sigma _{U\sigma }^{\mu \nu }(i\omega_n ) &=&UF_{\sigma }^{\mu \nu ^{\prime
}}(i\omega_n )\left[ {\hat G_{imp}}^{-1}\right] _{\sigma }^{\nu ^{\prime }\nu }(i\omega_n ) \\
\Sigma _{V\sigma }^{\mu \nu }(i\omega_n ) &=&K_{\sigma }^{\mu \nu ^{\prime
}}(i\omega_n )\left[ {\hat G_{imp}}^{-1}\right] _{\sigma }^{\nu ^{\prime }\nu }(i\omega_n )
\end{eqnarray}
and
\begin{eqnarray}
F_{\sigma }^{\mu \nu }&=& \ll c_{\mu \sigma }c_{\mu \bar{\sigma}
}^{\dagger }c_{\mu \bar{\sigma}},c_{\nu \sigma }^{\dagger }\gg \\
K_{\sigma }^{\mu \nu }&=& \sum_{\nu ^{\prime }}{V}_{\mu \nu
^{\prime }}\ll c_{\mu \sigma }n_{\nu ^{\prime }},c_{\nu \sigma }^{\dagger
}\gg \nonumber \\
&&+ \sum_{\nu ^{\prime }}{V}_{\nu ^{\prime }\mu }n_{\nu
^{\prime }}\ll c_{\mu \sigma },c_{\nu \sigma }^{\dagger }\gg
\end{eqnarray}
Here $\ll A,B \gg$ denotes the Green's function of operators $A$ and
$B$. This method of computing the self energy is particularly robust
for small values of $U/t$, when the self energy is relatively small in
magnitude, as compared to the direct approach.

\end{document}